\documentclass[twocolumn,showpacs,superscriptaddress]{revtex4}
\usepackage{amsfonts,amsthm,amsmath,amssymb,graphicx,hyperref,youngtab,calc}

\usepackage{hyperref}
\hypersetup{
colorlinks,breaklinks,
	urlcolor=[rgb]{0.,0.2,0.6},
	citecolor=[rgb]{0.7,0.1,0.1},
	linkcolor=[rgb]{0.,0.2,0.7}
}

\newcommand{\be}{\begin{equation}}
\newcommand{\ee}{\end{equation}}
\newcommand{\bea}{\begin{eqnarray}}
\newcommand{\eea}{\end{eqnarray}}

\newcommand{\nn}{\nonumber}
\newcommand{\Tr}{\text{Tr}}

\newcommand{\pd}{\partial}
\newcommand{\arctanh}{\text{arctanh}}

\begin{document}

\title{Gravitational anomalies, entanglement entropy, and flat-space holography}

\author{Seyed Morteza Hosseini}
\email{morteza.hosseini@mib.infn.it}
\affiliation{Dipartimento di Fisica, Universit\`a di Milano-Bicocca, I-20126 Milano, Italy}
\affiliation{INFN, sezione di Milano-Bicocca, I-20126 Milano, Italy}

\author{\'{A}lvaro V\'{e}liz-Osorio}
\email{alvaro.velizosorio@wits.ac.za}
\affiliation{Mandelstam Institute for Theoretical Physics, School of Physics\\University of the Witwatersrand, WITS 2050, Johannesburg, South Africa}

\date{\today}

\preprint{WITS-MITP-016}

\begin{abstract}

We introduce a prescription to compute the entanglement entropy of Galilean conformal field theories by combining gravitational anomalies and an \.{I}n\"{o}n\"{u}-Wigner contraction. We find that our expression for the entanglement entropy in the thermal limit reproduces the Cardy formula for Galilean conformal field theories.
Using this proposal, we calculate the entanglement entropy for a class of Galilean conformal field theories, which are believed to be dual to three-dimensional flat-space cosmological solutions.
These geometries describe expanding (contracting) universes and can be viewed as the flat-space limit of rotating Ba\~nados-Teitelboim-Zanelli black holes. We show that our finding reduces, in the appropriate limits, to the results discussed in the literature and provide interpretations for the previously unexplored regimes, such as flat-space chiral gravity.

\end{abstract}
\maketitle

\section{Introduction}
\label{intro}

Entanglement is the quintessential property of quantum systems.
The entanglement entropy (EE) is a measure of how much the different parts of a quantum system are entangled to each other. In \cite{Calabrese:2004eu}, it has been shown that the EE in (1+1)-dimensional conformal field theories (CFTs)
can be computed using the constraints imposed by conformal symmetries on the two-point functions, as well as the transformation properties of the stress-energy tensor. The problem of computing the EE for CFTs has been reformulated geometrically, in light of the holographic principle,
by Ryu and Takayanagi (RT) \cite{Ryu:2006bv}.
This insight has made clear that EE provides a valuable bridge between gravity and condensed matter physics.
Lately, the study of EE has been extended to CFTs where the left- and right- moving sectors
have different central charges \cite{Solodukhin:2005ah, Wall:2011kb, Castro:2014tta}.
These, correspond to theories that are sensitive to the coordinate system used to describe them, i.e., they exhibit gravitational anomalies.

Recently, for a class of nonrelativistic field theories which are governed by
the symmetries of the Galilean conformal algebra (GCA) the EE has been computed. Interestingly,
the GCA is isomorphic to the symmetry algebra of asymptotically flat spacetimes at null infinity, which
is called the (centrally extended)
Bondi-Metzner-Sachs (BMS)
algebra \cite{Bondi:1962px, Sachs:1962wk, Sachs:1962zza,Barnich:2010eb}.
This fact is at the root of the BMS/GCA correspondence
\cite{Bagchi:2009my,Bagchi:2010zz,Bagchi:2010eg,Bagchi:2012cy}, a flat-space analogue of the AdS/CFT duality \cite{Maldacena:1997re}.
This correspondence raises a number of compelling questions such as finding an analogue of the Cardy formula \cite{Bagchi:2012xr},
the flat-space stress-energy tensor \cite{Fareghbal:2013ifa} and the EE \cite{Bagchi:2014iea}, among others.
In this work, we pursue this line of inquiry to gain a deeper insight of the putative field theories dual to asymptotically flat spacetimes.

One can find asymptotically flat black holes in three dimensions only by including matter with negative energy;
technically, violating the dominant energy condition. Thus, it might seem that there are not any interesting (purely) gravitational configurations to study flat-space holography in (2+1) dimensions.
Nevertheless, there is a peculiar limit of the rotating 
Ba\~nados-Teitelboim-Zanelli
(BTZ) black hole \cite{Banados:1992wn}
 in which one can obtain a flat-space cosmology (FSC) geometry \cite{Cornalba:2002fi}.
This is a time-dependent solution of (2+1)-dimensional gravity theories, such as Einstein gravity and topologically massive gravity (TMG),
with a nontrivial Bekenstein-Hawking entropy associated to its cosmological horizon.
One of the main goals of this work is to compute the EE corresponding
to the (1+1)-dimensional Galilean conformal field theory ($\text{GCFT}_2$) dual to this spacetime. 

In order to find the EE for FSC, we propose a new approach that combines gravitational anomalies
and the $\dot {\rm I}$n$\ddot {\rm o}$n$\ddot {\rm u}$-Wigner contraction \footnote{In \cite{Krishnan:2013wta}, it has been been shown that interpreting the inverse AdS$_{3}$ radius $1/\ell$ as a Grassmann variable maps gravity in AdS$_{3}$ to gravity in flat space.}.
To test this approach, first we derive the EE for Galilean invariant field theories at zero temperature and find that this computation matches the result in \cite{Bagchi:2014iea}. Then we apply our method to calculate the EE of the FSC geometry. We find a general expression which reduces to known results in the appropriate limits and hints at new directions.

\section{Entanglement entropy in CFT$_2$ and Lorentz anomalies}

In quantum theories, even if we have access to all the information contained in a system, that does not guarantee that 
we can describe completely all of its subsystems. 
For each of these subsystems we can define a quantity called EE which quantifies this.
Imagine that the system of interest is described by a pure state $|\Psi\rangle$
and we wish to understand whether a subsystem $A$ is entangled with the rest.
The first step is to find the reduced density matrix, which is obtained by tracing out the degrees of freedom in the complement of $A$, $\bar A$
\be
\rho_A=\Tr_{{\cal H}_{\bar A}}|\Psi\rangle\langle \Psi|\, .
\ee
If there is any entanglement between the
degrees of freedom in $A$ and those in $\bar A$, to an observer having access only to $A$ the system appears to be in a mixed state.
If that is the case, then the von Neumann entropy of $\rho_A$
\be
S_A=-\Tr(\rho_A\log \rho_A)\, ,\label{eq:def EE}
\ee
is nonvanishing.
We refer to this quantity as the EE of $A$.
In practice, we use the so-called replica trick \cite{Calabrese:2004eu} and instead compute the R\'enyi entropies
\be
S_A^{(n)} =\frac{1}{1-n}\log\Tr\left[\rho^n_A\right]\,.\label{Renyi}
\ee
Then the EE \eqref{eq:def EE} can be extracted from \eqref{Renyi} by taking the $n\to 1$ limit. 
Before studying EE for a $\text{GCFT}_2$, let us briefly remind the reader how to compute this quantity for a $\text{CFT}_2$.
Suppose that the entangling region $A$ is a collection of $q$ disjoint intervals.
As shown in \cite{Calabrese:2004eu},  $S_A^{(n)}$  can be computed as a $2 q$-point function of twist operators $\Phi_n$ on the plane. For a $\text{CFT}_2$, these correspond to primary operators with scaling dimensions 
\be
  \Delta_n=\frac{c_L}{24}\left(n-\frac{1}{n}\right)\,,\qquad\bar\Delta_n=\frac{c_R}{24}\left(n-\frac{1}{n}\right)\,,
\ee
where $c_L$ ($ c_R$) is the central charge for the left- (right-) moving sector of the $\text{CFT}_2$.
Therefore, if $A$ corresponds to a single interval on the plane with end points $z_1$ and $z_2$ we have
\begin{align}\label{t dim}
\Tr\left(\rho_A^n\right)\propto
 \left(\frac{z_{12}}{\delta}\right)^{-2\Delta_n}\left(\frac{\bar z_{12}}{\delta}\right)^{-2\bar\Delta_n},
\end{align}
where $z_{12}=z_2-z_1$. Hereafter, without loss of generality we set $z_1=0$ and $z_2=z$.
 Using \eqref{eq:def EE}-\eqref{t dim}, the zero temperature EE for a single interval reads
\be
S_A=\left[{c_L\over 6} \log \Big({z\over \delta}\Big)
+{c_R\over 6} \log \Big({\bar{z}\over \delta}\Big)\right]\, .\label{EE CFT}
\ee
The separation of this quantity into a contribution coming from left movers and one from the right movers will be important for our future considerations.

In the following, we study the behavior of EE for a class of nonrelativistic CFTs.
In order to observe the asymmetry between space and time that arises for such theories we boost the entangling region.
Clearly, for Lorentz-invariant theories, the EE is expected to be invariant, however;
this is not the case for nonrelativistic CFTs. Strictly speaking, one investigates EE only for spacelike regions,
since EE measures entanglement between degrees of freedom and one pictures these as living in space,
and evolving through time. However, for the sake of the argument, let us rotate $z$ in the complex plane to $z = R e^{i\theta}$.
Under this rotation, the EE transforms into
\be
S_{\rm EE}= {c_L+c_R\over 6} \log \left({R\over \delta}\right)+ {c_L-c_R\over 6} i\theta\, .
\ee
Notice that the second term vanishes for theories where the Lorentz symmetry is nonanomalous (i.e., $c_L=c_R$) but is present otherwise \cite{Castro:2014tta}. In terms of spacetime variables, under the analytic continuation
$z=x-t,\bar z=x+t$, the angle $\theta$ corresponds to a boost parameter
$\kappa$ with $\theta=i\kappa$. Therefore, we find that the EE for the boosted interval reads
\be\label{eevac0}
S_{\rm EE}= {c_L+c_R\over 6} \log \left({R\over \delta}\right)- {c_L-c_R\over 6} \kappa\, .
\ee
Observe that this quantity is frame dependent for nonrelativistic CFTs.
In the next section we will focus on special nonrelativistic CFTs -- namely GCFTs.
\paragraph{Zero temperature EE for GCFT$_2$.}
The GCA is the infinite-dimensional algebra generated
by the conformal isometries of Galilean spacetimes.
In 1+1 dimensions, which we consider here,
the vector fields that generate these transformations are given by \cite{Bagchi:2012cy}
\begin{align}\label{GCAgen}
L_n=-(n+1)x^nt\pd_t-x^{n+1}\pd_x\, ,\qquad M_n=x^{n+1}\pd_t\, .
\end{align}
They satisfy the algebra
\begin{align}\label{GCA}
  \nonumber &[L_m,L_n]=(m-n)L_{n+m}+{c_{LL}\over 12}m(m^2-1)\delta_{m+n,0}\,,\\
  \nonumber &[L_m,M_n]=(m-n)M_{n+m}+{c_{LM}\over 12}m(m^2-1)\delta_{m+n,0}\, ,\\ 
            &[M_m,M_n]=0\, .
  \end{align}
In contrast, the (1+1)-dimensional relativistic conformal algebra is given by two copies of the Virasoro algebra
generated by ${\cal L}_n$ and $\bar {\cal L}_n$.
Notice that the GCA contains one copy of the Virasoro algebra and the mixed commutator has a very similar structure also. 
It is, therefore, natural to expect a relationship between these two algebras.
Indeed, it can be shown that the algebra \eqref{GCA} can be obtained
by performing an $\dot {\rm I}$n$\ddot {\rm o}$n$\ddot {\rm u}$-Wigner contraction of the relativistic conformal algebra \cite{Bagchi:2010zz,Bagchi:2010eg}.
This boils down to identifying the GCA generators with the linear combinations
	\be
	  L_n={\cal L}_n+\bar{\cal L}_n\, ,\quad
             M_n=-\epsilon\left({\cal L}_n-\bar{\cal L}_n\right)\, \label{IW}\, ,
	\ee
of their Virasoro counterparts, where $\epsilon$ is a small parameter at the level of algebra.
It is important to keep in mind that GCFTs are interesting on their own right,
and should not be regarded just as contractions of relativistic CFTs.
As such, the physical properties of GCFTs can be studied by considering the
constraints imposed by their symmetry algebra without any reference to a parent CFT. 

However, let us see whether the $\dot {\rm I}$n$\ddot {\rm o}$n$\ddot {\rm u}$-Wigner  contraction can provide a shortcut to compute the EE for GCFTs.
First we decompose $R$ into space and time as
\be
R^2=(\Delta x)^2-(\Delta t)^2\, ,\qquad
\kappa=\arctanh \left(\frac{\Delta t}{\Delta x}\right)\, .\label{deco}
\ee
Then we plug this back into \eqref{eevac0} and consider the $\dot {\rm I}$n$\ddot {\rm o}$n$\ddot {\rm u}$-Wigner contraction \eqref{IW}.
From a spacetime point of view, this corresponds to taking 
$\Delta t\to \epsilon \Delta t$ and $\Delta x\to  \Delta x$, in the $\epsilon\to 0$ limit.
This computation yields the EE for ${\rm GCFT}_{2}$, which reads
\be
S^{{\rm GCFT}_{2}}_{\rm EE}=\frac{c_{LL}}{6}\log \left(\frac{\Delta x}{\delta}\right)
+\frac{c_{LM}}{6}\left(\frac{\Delta t}{\Delta x}\right)\, ,\label{EE GCA}
\ee
and matches the result obtained in \cite{Bagchi:2014iea}. Let us close this section with a final comment:
The replica trick is valid for any quantum field theory; moreover, 
for (1+1)-dimensional field theories it implies that the EE of an interval is given by a two-point function of twist operators.
Now, if the theory is symmetric enough, it is  possible to constrain the functional form of the two-point functions.
For the case at hand, notice that in Eq.\,\eqref{EE GCA} the first contribution comes from the Virasoro sector of the theory,
while the second one comes from the exponential contribution to the two-point function found in \cite{Bagchi:2009ca}
employing the Ward identities corresponding to $M_n$. A similar behavior of the EE
is also exhibited by other nonrelativistic systems \cite{Hosseini:2015gua}.

\section{Flat-space holographic EE at zero temperature}

For (2+1)-dimensional asymptotically flat spacetimes, the symmetry algebra at null infinity is given by the semidirect sum of the infinitesimal diffeomorphisms on the circle with an Abelian ideal of supertranslations.
Moreover, it was shown that this algebra admits a nontrivial central extension.
In particular, for Einstein gravity it reads \cite{Barnich:2006av}
\begin{align}\label{BMS}
   &[{\cal J}_m,{\cal J}_n]=(m-n){\cal J}_{n+m}\,,\quad [{\cal P}_m,{\cal P}_n]=0\, , \\
   &[{\cal J}_m,{\cal P}_n]=(m-n){\cal P}_{n+m}+\frac{1}{4G}m(m^2-1)\delta_{m+n,0}\, . \nonumber
\end{align}
This is the so-called $\text{BMS}_3$ algebra corresponding to Einstein gravity.
Clearly, this algebra is isomorphic to a  $\text{GCA}_2$ \eqref{GCA} with $c_{LL}=0$ and $c_{LM}=3/G$,
where $G$ is the Newton constant
\footnote{There are two possible $\dot {\rm I}$n$\ddot {\rm o}$n$\ddot {\rm u}$-Wigner contractions
to get a GCFT$_2$ from a CFT$_2$, namely the nonrelativistic (NR) and the ultrarelativistic (UR) one.
Usually, the NR contraction is performed on the plane, while the UR contraction is done on a cylinder.
It is the latter that is of relevance for flat-space holography \cite{Bagchi:2012cy}.
Regardless of whether we use the NR or UR contraction we
obtain the GCFT$_2$ algebra and the symmetry can be used to constraint the form of the two-point functions.
Now, we can forget about the parent CFT and the contraction procedure.
Finally, the central charges can be read from the BMS$_3$ algebra, which comes directly from the asymptotic analysis.}.
Thus, invoking the BMS/GCA correspondence we see that
\eqref{EE GCA} reduces to 
\be
S^{{\rm GCFT}_{2}}_{\rm EE}=\frac{1 }{2 G }\left(\frac{\Delta t}{\Delta x}\right)\, ,
\ee
which can be interpreted as the holographic EE for (2+1)-dimensional asymptotically flat spacetimes at zero temperature \cite{Bagchi:2014iea}. Notice that this quantity vanishes whenever one computes EE for spacelike entangling regions.  
In the following, we will consider an analogous problem at finite temperature.

\section{Finite temperature EE for GCFT$_2$}
The EE for a CFT with gravitational anomaly ($c_L\neq c_R$) can be written as
\begin{align}\label{CFT EE FT}
S_{\text{EE}}=
\frac{c_{L}}{6}\log& \left[\frac{\beta_{L}}{\pi \delta }\sinh \left(\frac{\pi z}{\beta_{L} }\right)\right]\nn\\&+
\frac{c_{R}}{6}\log \left[\frac{\beta_{R}}{\pi \delta }\sinh \left(\frac{\pi \bar z}{\beta_{R} }\right)\right]\, ,
\end{align}
where $\beta_{L}$ ($\beta_{R}$) is the left- (right-) moving inverse temperature. Once again we want to break the Lorentz invariance of the theory; thus, we boost the interval and follow the same strategy used for the zero temperature case. This procedure yields
\begin{align}\label{EE GCFT FT}
 S^{\text{GCFT}_2}_{\text{EE}}&=\frac{c_{LL}}{6} \log \left[\frac{\beta _{LL} }{\pi  \delta }\sinh \left(\frac{\pi \Delta x}{\beta _{LL}}\right)\right]
-\frac{c_{LM}}{6}\frac{\beta _{LM}}{\beta _{LL}}\nn\\&
+\frac{ c_{LM} \pi}{6 \beta _{LL}}\left(\Delta t+\frac{\beta _{LM} }{\beta _{LL}}\Delta x\right)\coth \left(\frac{\pi  \Delta x}{\beta _{LL}}\right)\ ,
\end{align}
where we defined
\be
\beta_{LL}=\lim_{\epsilon\to 0}\frac{1}{2}(\beta_L+\beta_R)\, ,\qquad
\beta_{LM}=\lim_{\epsilon\to 0}\frac{1}{2\epsilon}(\beta_L-\beta_R)\, ,
\ee
which can be understood as the inverse temperatures of a GCFT$_2$.
Our expression for the EE yields the
GCFT Cardy formula in the thermal limit \footnote{To extract the thermal entropy we use the prescription $\Delta x\to\infty$:
\[
S_{{\rm EE}}\to \left(\frac{S_{{\rm thermal}}}{2\pi}\right)\Delta x+\dots \, .
\]} $\Delta x\gg (\beta_{LL},\beta_{LM})$, i.e.,
\be
S_{\rm thermal}^{{\rm GCFT}_2}=\frac{\pi^2}{3 \beta_{LL}^2}\left(c_{LL}\beta_{LL} + c_{LM}\beta_{LM}\right)\, .
\ee
Observe that as we take $c_{LM}=0$ the above expression reduces to the Cardy formula for a chiral CFT$_2$. This is consistent with the fact that the 
GCA\,\eqref{GCA} contains a Virasoro subalgebra with central charge $c_{LL}$.
In the next section we will show that the above result can be interpreted as the holographic EE for FSC.

\section{Flat-space holographic EE at finite temperature}

We saw above, that the GCFT dual to Einstein gravity has only one nonvanishing central charge.
Hence, part of the structure of the EE cannot be seen in that context.
To be able to see the full structure of the holographic EE we consider a theory for which both
$c_{LL}$ and $c_{LM}$ are nonzero, namely topological massive gravity \cite{Deser:1982vy} whose action reads
\be\label{TMG}
I_{\rm TMG}=\frac{1}{16\pi G}\int d^{3}x \sqrt{-g}\left[R-2\Lambda
+\frac{1}{2\mu}{\rm CS}(\Gamma)\right],
\ee
where the Chern-Simons contribution is given by 
\be
{\rm CS}(\Gamma)=\varepsilon ^{\alpha \beta \gamma}\Gamma _{\;\alpha \sigma }^{\rho }
\left(\partial _{\beta }\Gamma _{\;\gamma \rho}^{\sigma }
+\frac{2}{3}\Gamma _{\;\beta \eta }^{\sigma }\Gamma _{\;\gamma \rho}^{\eta }\right)\, .
\ee
Canonical analysis of TMG yields a BMS$_3$ algebra with \cite{Bagchi:2012yk}
\be
c_{LL}=\frac{3}{G \mu }\, ,\qquad
c_{LM}=\frac{3}{G}\, .\label{TMG cc}
\ee
Notice that the Einstein gravity limit corresponds to taking $\mu\to\infty$.

The equations of motion of TMG admit a rotating BTZ solution, whose metric can be written as
\begin{align}\label{metricbtz}
\frac{ds_{\rm BTZ}^2}{\ell^2}&=-\frac{(r^2 - r_+^2)(r^2 - r_-^2)}{r^2} d\tau^2\nonumber\\&
+\frac{r^2}{(r^2 - r_+^2)(r^2 - r_-^2)} dr^2
+ r^2 \left(d\phi + \frac{r_+ r_-}{r^2}d\tau\right)^2 \ ,
\end{align}
where $\ell$ is the AdS radius of curvature.
We define the quantities
\be
m=\frac{r_+^2-r_-^2}{8 G \ell^2}\, ,\qquad j=\frac{r_+ r_-}{4 G \ell}\, .
\ee
In pure Einstein gravity $m$ and $j$ correspond to the mass and angular momentum of the black hole. Due to the Chern-Simons term in the 
action there is a shift in the conserved quantities associated to the TMG-BTZ black hole
\be
M=m-\frac{j}{\mu\ell^2}\, ,\qquad J=j-\frac{m}{\mu}\, .
\ee
This spacetime is dual to a field theory state with left- and right-moving asymmetric ensembles with inverse temperatures
\be
\beta_L = \frac{2\pi \ell}{r_+ - r_-}\, , \qquad \beta_R = \frac{2\pi \ell}{r_+ + r_-} \ .
\ee
In the following, we will unwrap the angular coordinate $\phi$. Hence, strictly speaking,
we are actually studying a boosted black brane that is a rotating BTZ black hole at high temperature.

The holographic EE for the geometry \eqref{metricbtz} in TMG is given by
\begin{align}
S_{\rm EE}^{\rm BTZ}&=\frac{c_L + c_R}{12}\log\left[\frac{\beta_L \beta_R}{\pi^2 \delta^2}
 \sinh\left(\frac{\pi z}{\beta_L}\right) \sinh\left(\frac{\pi \bar z}{\beta_R}\right)\right]\nonumber\\&
-\frac{c_L - c_R }{12} \log\left[\frac{\sinh\left(\frac{\pi \bar z}{\beta_R}\right) \beta_R}
{\sinh\left(\frac{\pi z}{\beta_L}\right) \beta_L}\right].\label{entEE}
\end{align}
The first term of the above equation was found for Einstein gravity in \cite{Hubeny:2007xt},
while the contribution due to the Chern-Simons term was computed more recently in \cite{Castro:2014tta}.
Of course, Eq.\,\eqref{entEE} is a suggestive form of writing Eq.\,\eqref{CFT EE FT} where one can 
explicitly see the term coming from the gravitational anomaly, which is proportional to $c_L-c_R$.

Now, we turn to the study of the flat-space analogue of the rotating BTZ black hole solution.
The flat-space limit corresponds to taking
the cosmological constant $\Lambda$ to zero; equivalently, we can consider the $\ell\to \infty$ limit.
In this limit, \eqref{metricbtz} describes an asymptotically flat expanding universe with line element \cite{Cornalba:2002fi}
\begin{align}
ds_{\rm FSC}^2&=\hat r_{+}^{2}\left(1-\frac{r_{0}^2}{r^2}\right)d\tau^{2}
-\frac{dr^2}{\hat r_{+}^{2}\left(1-\frac{r_{0}^2}{r^2}\right)}\nonumber\\&
+r^2\left(d\phi-\frac{\hat r_{+}r_{0}}{r^2}d\tau\right)^2\, ,
\end{align}
where 
\be
r_+\to\ell\sqrt{8Gm}=\ell\hat r_{+}\, ,\qquad r_-\to\sqrt{\frac{2G}{m}}|j|=r_{0}\, .\label{para}
\ee
This metric can be seen as a shifted-boost orbifold of Minkowski spacetime.
Hereafter, we refer to this geometry as the FSC.
After taking the flat-space limit, the outer horizon of BTZ is mapped to infinity,
while the inner horizon remains at a finite radius and defines a cosmological horizon. 
One can associate an entropy and a temperature to this horizon.
In \cite{Bagchi:2012xr}, the authors derived the Bekenstein-Hawking entropy of FSC in Einstein gravity by counting $\text{GCFT}_2$ microstates.

Finally, we bring the pieces together to compute the holographic EE for FSC.
Since the FSC is the flat-space limit of the rotating BTZ black hole,
we claim that its holographic EE can be found by following our prescription to implement an $\dot {\rm I}$n$\ddot {\rm o}$n$\ddot {\rm u}$-Wigner contraction on the holographic EE of the rotating BTZ black hole \eqref{entEE}. More explicitly, we write $z = R e^{i\theta}$, which, just as before, is related to the boost parameter by $\theta=i\kappa$. Then, employing Eq.\,\eqref{deco} and taking the $\epsilon\to 0$ limit, we find
\begin{align}\label{EE FSC}
 S^{\text{FSC}}_{\text{EE}}&=\frac{c_{LL}}{6} \log \left[\frac{\beta _+ }{\pi  \delta }\sinh \left(\frac{\pi \Delta x}{\beta _+}\right)\right]
-\frac{c_{LM}}{6}\frac{\beta _0}{\beta _+}\nn\\&
+\frac{ c_{LM} \pi}{6 \beta _+}\left(\Delta t+\frac{\beta _0 }{\beta _+}\Delta x\right)\coth \left(\frac{\pi  \Delta x}{\beta _+}\right)\ ,
\end{align}
where 
\be
\beta_{+}=\frac{2 \pi }{\hat{r}_+}\, , \qquad
\beta_{0}=\frac{2 \pi  r_0}{\hat{r}_+^2}\, .\label{temps}
\ee
This is one of the main results in this paper, which is in agreement with our intrinsic GCFT computation \eqref{EE GCFT FT}, with $\beta_+=\beta_{LL}$ and $\beta_0=\beta_{LM}$. Thus, Eq.\,\eqref{EE FSC} provides a holographic realization of Eq.\,\eqref{EE GCFT FT}, where the AdS$_3$ radius has been identified with the inverse of the $\dot {\rm I}$n$\ddot {\rm o}$n$\ddot {\rm u}$-Wigner parameter (which implies a vanishing cosmological constant).

Notice that in the zero temperature limit, i.e., $\beta_+\to\infty$,
Eq.\,\eqref{EE FSC} reduces to Eq.\,\eqref{EE GCA}.
Recall that for any (1+1)-dimensional quantum field theory, whenever the entangling region corresponds to a single interval,
the R\'enyi entropies can be determined by computing a two-point function of twist operators.
Just as for $\text{CFT}_2$, for a $\text{GCFT}_2$ the functional form of these two-point functions is completely fixed by symmetry.
The next step is to find the quantum numbers of the twist operators;
for a $\text{CFT}_2$ these are the weights $(\Delta_n,\bar\Delta_n)$ and for a  $\text{GCFT}_2$
they correspond to the weight with respect to the Virasoro sector and the rapidity associated
with the $M_n$ generators (see \cite{Bagchi:2009ca} for more details).
For a  $\text{GCFT}_2$ these quantum numbers are determined by the central charges of the GCA, $c_{LL}$ and $c_{LM}$ in Eq.\,\eqref{TMG cc}.
It is these quantities that we import from the asymptotic analysis via the BMS/GCA correspondence.

\section{Interesting limits of the FSC holographic EE}
In the present section, we consider some relevant limits of formula \eqref{EE FSC}, which will serve both as consistency checks and as invitations for further research.
\begin{enumerate}
\item \textit{Einstein gravity.---} From the action \eqref{TMG} it can be seen that we recover Einstein gravity by considering the limit $\mu\to\infty$.
This limit corresponds to setting $c_{LL}=0$ in Eq.\,\eqref{EE FSC} as can be seen from Eq.\,\eqref{TMG cc}.
Therefore, as $\mu\to\infty$ we find
\be
\hspace{7mm} \frac{c_{LM}}{6\beta _+}\left[\pi \left(\Delta t+\frac{\beta _0 }{\beta _+}\Delta x\right)\coth \left(\frac{\pi  \Delta x}{\beta _+}\right)
-\beta _0\right]\, .
\ee
Notice that even if we consider a spacelike entangling region, there is a residual EE proportional to the inverse temperature of the cosmological horizon, $\beta_0$.

\item\textit{Boost orbifold.---} 
This limit corresponds to taking $j$ to zero;
combining \eqref{para} and \eqref{temps} it is clear that this amounts to setting $\beta_0=0$ in Eq.\,\eqref{EE FSC}.
This is the regime where
there is no cosmological horizon to hide the singularity and can be regarded as a big bang toy model.
Moreover, it is apparent that the corresponding geometry is closely related to the Rindler spacetime via Wick rotation.
For this case, the EE reads
\begin{align}
\frac{c_{LL}}{6} \log &\left[\frac{\beta _+ }{\pi  \delta }\sinh \left(\frac{\pi \Delta x}{\beta _+}\right)\right]\nn\\&
+\frac{ c_{LM} \pi}{6 \beta _+}\Delta t\coth \left(\frac{\pi  \Delta x}{\beta _+}\right)\, ,
\end{align}
which matches the GCFT finite temperature result of \cite{Bagchi:2014iea}.
It would be quite interesting to explore this line of research since nonextremal
black holes contain a universal Rindler patch in their near horizon region.
Perhaps this result can shed some light on the understanding of the Bekenstein-Hawking entropy of nonextremal black holes.

\item \textit{Thermal entropy.---} The thermal entropy of the system can be recovered by considering a large entangling region.
Hence, in this limit ($\Delta x\to\infty$) we expect to find the Bekenstein-Hawking entropy for FSC in TMG.
Applying this limit to \eqref{EE FSC}, we find
\be
S^{\text{FSC}}_{\text{thermal}}=\frac{2\pi r_{0}}{4G}+\frac{2\pi \hat r_{+}}{4 G \mu}\, ,
\ee
where we have used the Eqs.\,\eqref{TMG cc} and \eqref{temps}. This expression is in agreement with the result previously reported in \cite{ Bagchi:2013qva}.

\item\textit{Flat-space chiral gravity.---} Finally, we consider the case where the bulk theory consists only of the gravitational Chern-Simons piece. 
One can obtain conformal Chern-Simons gravity (CSG) by scaling Newton\rq{}s constant to infinity,
while keeping $\mu G=1/8k$ fixed \cite{Bagchi:2012yk}. The CSG action reads
\be\label{CSG}
I_{\rm CSG}=\frac{k}{4\pi}\int d^{3}x \sqrt{-g}\,{\rm CS}(\Gamma)\, ,
\ee
where the constant $k$ is the Chern-Simons level.
In this limit we see that $c_{LL}=24k $ and $c_{LM}=0$; therefore, Eq.\,\eqref{EE FSC} implies that the EE of FSC becomes
\be
S^{\chi}_{\text{EE}}=
\frac{c_{LL}}{6}\log \left[\frac{\beta_{+}}{\pi \delta }\sinh \left(\frac{\pi  \Delta x}{\beta_{+} }\right)\right]\, .
\ee
Interestingly, our result for the CSG EE has exactly the form of a chiral CFT at finite temperature,
this is consistent with the fact that for CSG the asymptotic symmetry algebra reduces to a chiral copy of the Virasoro algebra.
\end{enumerate}

Thus, we have found that the FSC EE formula \eqref{EE FSC} captures many different scenarios. In the familiar ones, it reduces to the correct results presented in the literature, while in those hitherto unexplored it predicts interesting behaviors that deserve further investigation. 

\paragraph{Discussion and outlook.} 
In this paper, we have advocated that using the $\dot {\rm I}$n$\ddot {\rm o}$n$\ddot {\rm u}$-Wigner contraction in combination with gravitational
anomalies allows one to compute the EE for field theories dual to asymptotically flat spacetimes in a succinct manner.
It would be interesting to understand the geometrical derivation of these results in terms of the RT prescription \cite{Ryu:2006bv}.

One can find asymptotically flat black holes in 2+1 dimensions by considering higher-derivative gravity theories.
Some properties of these black holes have been studied employing the BMS/GCA correspondence \cite{Fareghbal:2014kfa}.
On the other hand, in the presence of higher curvature corrections, the holographic EE computations encounter many subtleties (see, e.g., \cite{Hosseini:2015vya}).
It is an interesting future problem to study the implications of the BMS/GCA correspondence
to investigate the holographic EE for these geometries.

Recently, there has been much interest in studying the changes in entanglement due to excitations in the system \cite{Nozaki:2014hna,Alcaraz:2011tn}.
Moreover, it has been found that these changes encode physically interesting quantities (see, e.g., \cite{He:2014mwa,Caputa:2015tua}).
One of the intriguing future directions is to try to address these questions for Galilean invariant field theories.
The formalism developed in this paper can be easily adapted to deal with these questions.

\section{Acknowledgments}
We are grateful to Reza Fareghbal and Kevin Goldstein for commenting on a draft of this paper.
We would like to thank Arpan Bhattacharyya, Pasquale Calabrese, Joan Camps,
and Shajid Haque for illuminating conversations and correspondence. Additionally, we wish to acknowledge Daniel Grumiller for comments on an earlier version of this work. 
The research of A.V.-O is supported by the University Research Council of the University of the Witwatersrand.
The work of S.M.H is supported in part by INFN. S.M.H and A.V.-O wish to thank the hospitality of the ICTP
in Trieste where part of this research was conducted. A.V.-O wishes to thank the visitors program of the HECAP section at ICTP
for their support.


\begin{thebibliography}{99}

\bibitem{Calabrese:2004eu} 
  P.~Calabrese and J.~L.~Cardy,
  `Entanglement entropy and quantum field theory,''
  J.\ Stat.\ Mech.\  {\bf 0406}, P06002 (2004)
  [hep-th/0405152].

\bibitem{Ryu:2006bv} 
  S.~Ryu and T.~Takayanagi,
  ``Holographic derivation of entanglement entropy from AdS/CFT,''
  Phys.\ Rev.\ Lett.\  {\bf 96}, 181602 (2006)
  [hep-th/0603001].

\bibitem{Solodukhin:2005ah} 
  S.~N.~Solodukhin,
``Holography with gravitational Chern-Simons,''
  Phys.\ Rev.\ D {\bf 74}, 024015 (2006)
  [hep-th/0509148].

\bibitem{Castro:2014tta} 
  A.~Castro, S.~Detournay, N.~Iqbal and E.~Perlmutter,
  ``Holographic entanglement entropy and gravitational anomalies,''
  JHEP {\bf 1407}, 114 (2014)
  [arXiv:1405.2792 [hep-th]].

\bibitem{Wall:2011kb} 
  A.~C.~Wall,
  ``Testing the Generalized Second Law in 1+1 dimensional Conformal Vacua: An Argument for the Causal Horizon,''
  Phys.\ Rev.\ D {\bf 85}, 024015 (2012)
  [arXiv:1105.3520 [gr-qc]].

\bibitem{Bondi:1962px} 
  H.~Bondi, M.~G.~J.~van der Burg and A.~W.~K.~Metzner,
  ``Gravitational waves in general relativity. 7. Waves from axisymmetric isolated systems,''
  Proc.\ Roy.\ Soc.\ Lond.\ A {\bf 269}, 21 (1962).

\bibitem{Sachs:1962wk} 
  R.~K.~Sachs,
  ``Gravitational waves in general relativity. 8. Waves in asymptotically flat space-times,''
  Proc.\ Roy.\ Soc.\ Lond.\ A {\bf 270}, 103 (1962).

\bibitem{Sachs:1962zza} 
  R.~Sachs,
  ``Asymptotic symmetries in gravitational theory,''
  Phys.\ Rev.\  {\bf 128}, 2851 (1962).

\bibitem{Barnich:2010eb} 
  G.~Barnich and C.~Troessaert,
  ``Aspects of the BMS/CFT correspondence,''
  JHEP {\bf 1005}, 062 (2010)
  [arXiv:1001.1541 [hep-th]].

\bibitem{Bagchi:2009my} 
  A.~Bagchi and R.~Gopakumar,
  ``Galilean Conformal Algebras and AdS/CFT,''
  JHEP {\bf 0907}, 037 (2009)
  [arXiv:0902.1385 [hep-th]].

\bibitem{Bagchi:2010zz} 
  A.~Bagchi,
  ``Correspondence between Asymptotically Flat Spacetimes and Nonrelativistic Conformal Field Theories,''
  Phys.\ Rev.\ Lett.\  {\bf 105}, 171601 (2010).

\bibitem{Bagchi:2010eg} 
  A.~Bagchi,
  ``The BMS/GCA correspondence,''
  arXiv:1006.3354 [hep-th].

\bibitem{Bagchi:2012cy} 
  A.~Bagchi and R.~Fareghbal,
  ``BMS/GCA Redux: Towards Flat Space Holography from Non-Relativistic Symmetries,''
  JHEP {\bf 1210}, 092 (2012)
  [arXiv:1203.5795 [hep-th]].

\bibitem{Maldacena:1997re} 
  J.~M.~Maldacena,
  ``The Large N limit of superconformal field theories and supergravity,''
  Int.\ J.\ Theor.\ Phys.\  {\bf 38}, 1113 (1999)
  [Adv.\ Theor.\ Math.\ Phys.\  {\bf 2}, 231 (1998)]
  [hep-th/9711200].

\bibitem{Bagchi:2012xr} 
  A.~Bagchi, S.~Detournay, R.~Fareghbal and J.~Simón,
  ``Holography of 3D Flat Cosmological Horizons,''
  Phys.\ Rev.\ Lett.\  {\bf 110}, no. 14, 141302 (2013)
  [arXiv:1208.4372 [hep-th]].

\bibitem{Fareghbal:2013ifa} 
  R.~Fareghbal and A.~Naseh,
  ``Flat-Space Energy-Momentum Tensor from BMS/GCA Correspondence,''
  JHEP {\bf 1403}, 005 (2014)
  [arXiv:1312.2109 [hep-th]].

\bibitem{Bagchi:2014iea} 
  A.~Bagchi, R.~Basu, D.~Grumiller and M.~Riegler,
  ``Entanglement entropy in Galilean conformal field theories and flat holography,''
  Phys.\ Rev.\ Lett.\  {\bf 114}, no. 11, 111602 (2015)
  [arXiv:1410.4089 [hep-th]].

\bibitem{Banados:1992wn} 
  M.~Banados, C.~Teitelboim and J.~Zanelli,
``The Black hole in three-dimensional space-time,''
  Phys.\ Rev.\ Lett.\  {\bf 69}, 1849 (1992)
  [hep-th/9204099].

\bibitem{Cornalba:2002fi}
  L.~Cornalba and M.~S.~Costa,
  ``A New cosmological scenario in string theory,''
  Phys.\ Rev.\ D {\bf 66} (2002) 066001
  [hep-th/0203031].

\bibitem{Bagchi:2009ca} 
  A.~Bagchi and I.~Mandal,
  ``On Representations and Correlation Functions of Galilean Conformal Algebras,''
  Phys.\ Lett.\ B {\bf 675}, 393 (2009)
  [arXiv:0903.4524 [hep-th]].

\bibitem{Hosseini:2015gua} 
  S.~M.~Hosseini and A.~Veliz-Osorio,
  ``Entanglement and mutual information in two-dimensional nonrelativistic field theories,''
  Phys.\ Rev.\ D {\bf 93}, no. 2, 026010 (2016)
  [arXiv:1510.03876 [hep-th]].

\bibitem{Barnich:2006av} 
  G.~Barnich and G.~Compere,
  ``Classical central extension for asymptotic symmetries at null infinity in three spacetime dimensions,''
  Class.\ Quant.\ Grav.\  {\bf 24}, F15 (2007)
  [gr-qc/0610130].

\bibitem{Deser:1982vy} 
  S.~Deser, R.~Jackiw and S.~Templeton,
  ``Three-Dimensional Massive Gauge Theories,''
  Phys.\ Rev.\ Lett.\  {\bf 48}, 975 (1982).

\bibitem{Bagchi:2012yk} 
  A.~Bagchi, S.~Detournay and D.~Grumiller,
  ``Flat-Space Chiral Gravity,''
  Phys.\ Rev.\ Lett.\  {\bf 109}, 151301 (2012)
  [arXiv:1208.1658 [hep-th]].

\bibitem{Hubeny:2007xt} 
  V.~E.~Hubeny, M.~Rangamani and T.~Takayanagi,
  ``A Covariant holographic entanglement entropy proposal,''
  JHEP {\bf 0707}, 062 (2007)
  [arXiv:0705.0016 [hep-th]].

\bibitem{Bagchi:2013qva} 
  A.~Bagchi and R.~Basu,
  ``3D Flat Holography: Entropy and Logarithmic Corrections,''
  JHEP {\bf 1403}, 020 (2014)
  [arXiv:1312.5748 [hep-th]].

\bibitem{Hosseini:2015vya} 
  S.~M.~Hosseini and A.~Veliz-Osorio,
  ``Free-kick condition for entanglement entropy in higher curvature gravity,''
  arXiv:1505.00826 [hep-th].

\bibitem{Fareghbal:2014kfa} 
  R.~Fareghbal and S.~M.~Hosseini,
  ``Holography of 3D Asymptotically Flat Black Holes,''
  Phys.\ Rev.\ D {\bf 91}, no. 8, 084025 (2015)
  [arXiv:1412.2569 [hep-th]].

\bibitem{Nozaki:2014hna}
  M.~Nozaki, T.~Numasawa and T.~Takayanagi,
  ``Quantum Entanglement of Local Operators in Conformal Field Theories,''
  Phys.\ Rev.\ Lett.\  {\bf 112} (2014) 111602
  [arXiv:1401.0539 [hep-th]].

\bibitem{Alcaraz:2011tn} 
  F.~C.~Alcaraz, M.~I.~Berganza and G.~Sierra,
  ``Entanglement of low-energy excitations in Conformal Field Theory,''
  Phys.\ Rev.\ Lett.\  {\bf 106}, 201601 (2011)
  [arXiv:1101.2881 [cond-mat.stat-mech]].

\bibitem{Caputa:2015tua}
  P.~Caputa and A.~Veliz-Osorio,
  ``Entanglement constant for conformal families,''
  Phys.\ Rev.\ D {\bf 92} (2015) 6,  065010
  [arXiv:1507.00582 [hep-th]].

\bibitem{He:2014mwa}
  S.~He, T.~Numasawa, T.~Takayanagi and K.~Watanabe,
  ``Quantum dimension as entanglement entropy in two dimensional conformal field theories,''
  Phys.\ Rev.\ D {\bf 90} (2014) 4,  041701
  [arXiv:1403.0702 [hep-th]].

\bibitem{Krishnan:2013wta} 
  C.~Krishnan, A.~Raju and S.~Roy,
  ``A Grassmann path from $AdS_3$ to flat space,''
  JHEP {\bf 1403}, 036 (2014)
  [arXiv:1312.2941 [hep-th]].

\end{thebibliography}
\end{document}